\documentclass[aps,prl,reprint,groupedaddress,showpacs,amsmath,amssymb,twocolumn,floatfix,preview]{revtex4-1}
\usepackage{graphicx}
\usepackage{bm}
\usepackage{color}
\usepackage[normalem]{ulem}
\usepackage{cleveref}

\begin{document}

\title{Even denominator fractional quantum Hall state in bilayer graphene}

\author{J.I.A. Li$^{1}$}
\author{C. Tan$^{2}$}
\author{S. Chen$^{3}$}
\author{Y. Zeng$^{1}$}
\author{T. Taniguchi$^{4}$}
\author{K. Watanabe$^{4}$}
\author{J. Hone$^{2}$}
\author{C.R. Dean$^{1}$}

\affiliation{$^{1}$Department of Physics, Columbia University, New York, NY, USA}
\affiliation{$^{3}$Department of Applied Physics and Applied Mathematics, Columbia University, New York, NY, USA}
\affiliation{$^{2}$Department of Mechanical Engineering, Columbia University, New York, NY, USA}
\affiliation{$^{4}$National Institute for Materials Science, 1-1 Namiki, Tsukuba, Japan}

\date{\today}




\maketitle

\textbf{The multi-component nature of bilayer graphene (BLG), together with the ability to controllably tune between the various ground state orders, makes it a rich system in which to explore interaction driven phenomena ~\cite{Mccann.06,Shibata.09,Chak.10,Sni.12}.  In the fractional quantum Hall effect (FQHE) regime, the unique Landau level spectrum of BLG is anticipated to support a  non-Abelian even-denominator state that is tunable by both electric and magnetic fields~\cite{Pap.11,Chak.11,Pap.14}.  However,  observation of this state, which is anticipated to be stronger than in conventional systems, has been conspicuously difficult ~\cite{Mor.14,Smet.15,Zib.16}. Here we report transport measurements of a robust even denominator FQHE in high-mobility, dual gated BLG devices.  We confirm that the stability of the energy gap can be sensitively tuned and map the phase diagram. Our results establish BLG as a dynamic new platform to study topological ground states with possible non-Abelian excitations. }

Since the discovery of the even denominator fractional quantum Hall (FQH) state, appearing at $5/2$ filling in the $N = 1$ Landau level (LL) of GaAs~\cite{Wil.87,Pan.99}, it has remained an intensely studied anomaly in condensed matter physics. 
Numerical calculations  suggest  the likely ground state to be the Moore-Read (MR) Pfaffian-a type of p-wave superconductor resulting from the condensation of composite Fermion pairs ~\cite{Morf.98,Moo.91,Gre.91}. The ground state at half filling is sensitive to the details of Coulomb interaction, and the $N=1$ Landau level of GaAs proved to be optimal for stabilizing the Pfaffian state ~\cite{Rezayi.00}. Experimental studies indicate that the $5/2$ state is spin polarized,  and carries $e/4$ fractional charge ~\cite{Willett.13}, observations which are both consistent with the Pfaffian description.  However, definitive confirmation remains elusive.  One of the most uniquely identifiable features of the Pfaffian is that its excitations are anticipated to be Majorana zero modes, obeying non-Abelian fractional statistics ~\cite{Moo.91,Read.96,Nayak.08}. In recent years this has gained increased attention with the growing interest in exploiting such exotic topological states for fault tolerant quantum computation ~\cite{Kit.03, Nayak.08}.  However, measurement of the quantum statistics associated with the $5/2$ state remains an open challenge, with progress hindered by the fragile nature of the $5/2$ state combined with unfavorable electrostatics in GaAs ~\cite{Wil.09,Halperin.11,Bishara.08}.

Bilayer graphene (BLG) encapsulated by hexagonal boron nitride (hBN)  offers a promising new platform for studying even denominator states. The $N=1$ LL in BLG has an unusual composition, comprising a mixture of the conventional Landau orbital $0$ and $1$ wavefunctions~\cite{Chak.11,Hunt.16}, and moreover is accidentally degenerate with the $N=0$ LL~\cite{Mccann.06} (Fig.~1a).  Due to this construction,  application of a strong magnetic field, $B$, can both lift the level degeneracy and modify the precise structure of the $N=1$ wavefunction by modifying the relative weight of the conventional $0,1$ contributions. Both of these effects determine the stability of the Pfaffian ground sate ~\cite{Pap.11,Chak.11}.  
Additionally, applying a transverse electric field, $D$, breaks the inversion symmetry between the two sets of graphene lattices, and induces phase transitions between ground states with different valley and orbital polarizations ~\cite{Yacoby.10,Hunt.16}. Application of both $B$ and $D$ fields therefore  makes it theoretically possible to dynamically tune several key parameters within a single device, including the orbital wavefunction, effective Coulomb interaction, and Landau level mixing, all of which play critical roles in determining the nature of the even denominator state ~\cite{Pap.11,Chak.11,Pap.14}. 

Here we report robust magneto-transport signatures of the even-denominator FQHE in ultra-high mobility BN-encapsulated BLG.  Utilizing a dual-gate geometry (Fig.~1b), to tune through different orbital and layer polarizations, we find four even denominator states appearing only within the symmetry broken  $N=1$ orbital branches of the lowest LL. Such orbital selection rule is consistent with theoretical expectations for the MR Pfaffian in BLG~\cite{Pap.11,Chak.11,Pap.14}. We investigate how these states evolve with varying  $B$ and $D$,  and provide the first  mapping of the $B-D$ phase diagram.  We confirm the unique tunability of the even denominator states in BLG, and reach a regime where the energy gap is found to exceed $1$~K.

\begin{figure*}
\includegraphics[width=1\linewidth]{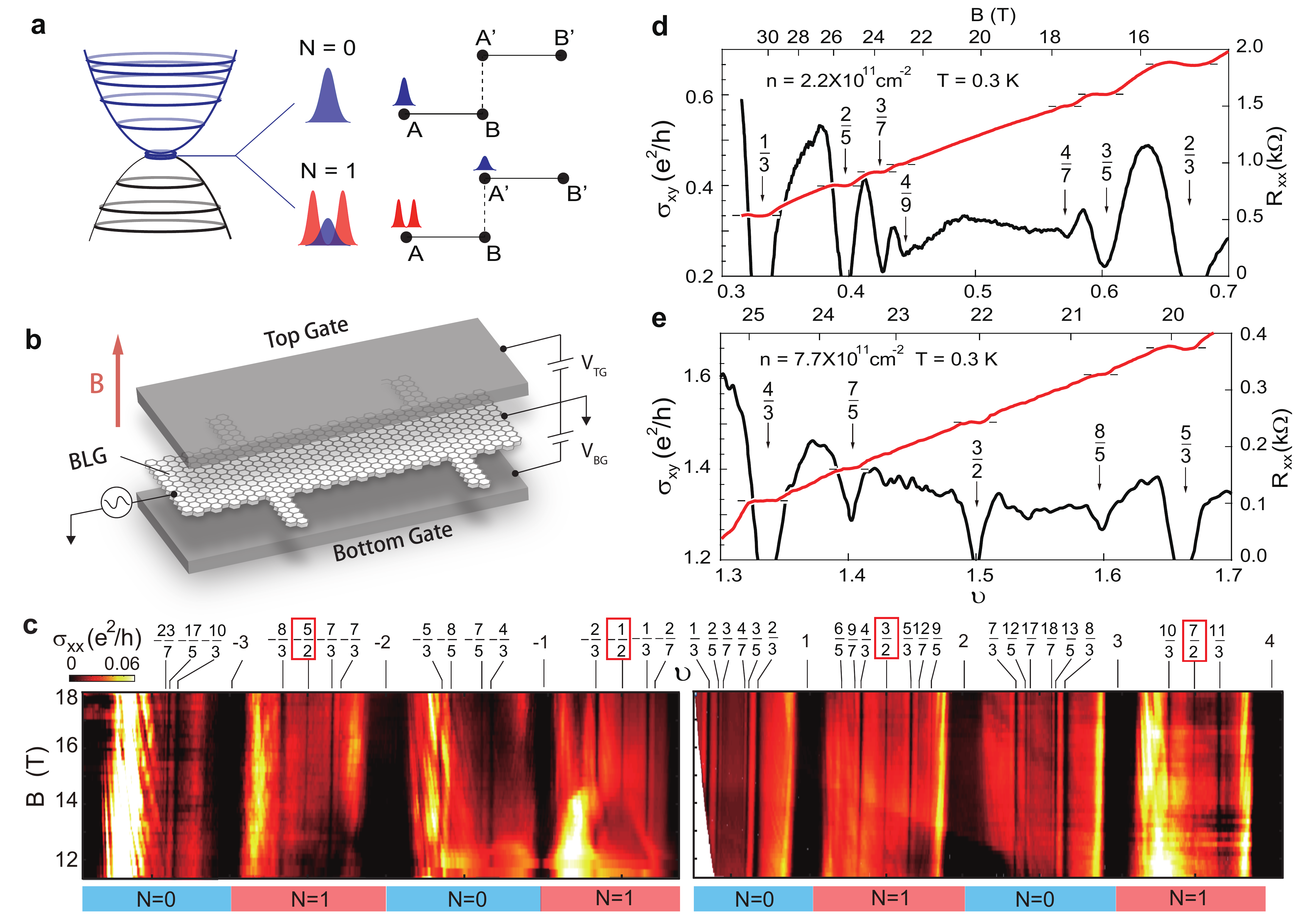}
\caption{\label{fig1}   (a) Energy spectrum of the BLG LLs near the charge neutrality point (CNP). The lowest two LLs with Landau orbital index $N=0$ and $1$ are energetically degenerate, leading to the eight-fold degeneracy of the LLL. The BLG wavefunction with Landau orbital $N=0$ is identical to the lowest Landau orbital wavefunction for conventional (non-relativistic) systems, whereas the orbital $N=1$ wavefunction is a mixture of the conventional Landau orbitals $0$ and $1$. The schematic shows the wavefunction distribution on the four atomic sites of BLG, for the $N=0$ and $1$ Landau orbital states.  (b) Schematic of the device geometry.  (c) $\sigma_{xx}$ as a function of filling factor $\nu$ and magnetic field $B$ at $T = 0.2$ K and (left) $D = -100$ mV/nm, (right) $D = 35$ mV/nm  for the LLL ($-4 \leq \nu \leq 4$). (d) and (e) show $\sigma_{xx}$ and $R_{xy}$, acquired by sweeping B at fixed carrier densities,  $n = 2.2 \times 10^{11}$ cm$^{-2}$ and $7.7 \times 10^{11}$ cm$^{-2}$, corresponding to filling fractions spanning $0 \leq \nu \leq 1$, and $1 \leq \nu \leq 2$, respectively.   Bottom axis labels the filling fraction $\nu$, with corresponding $B$ values on the top axis.      }
\end{figure*}

\begin{figure*}
\includegraphics[width=0.9\linewidth]{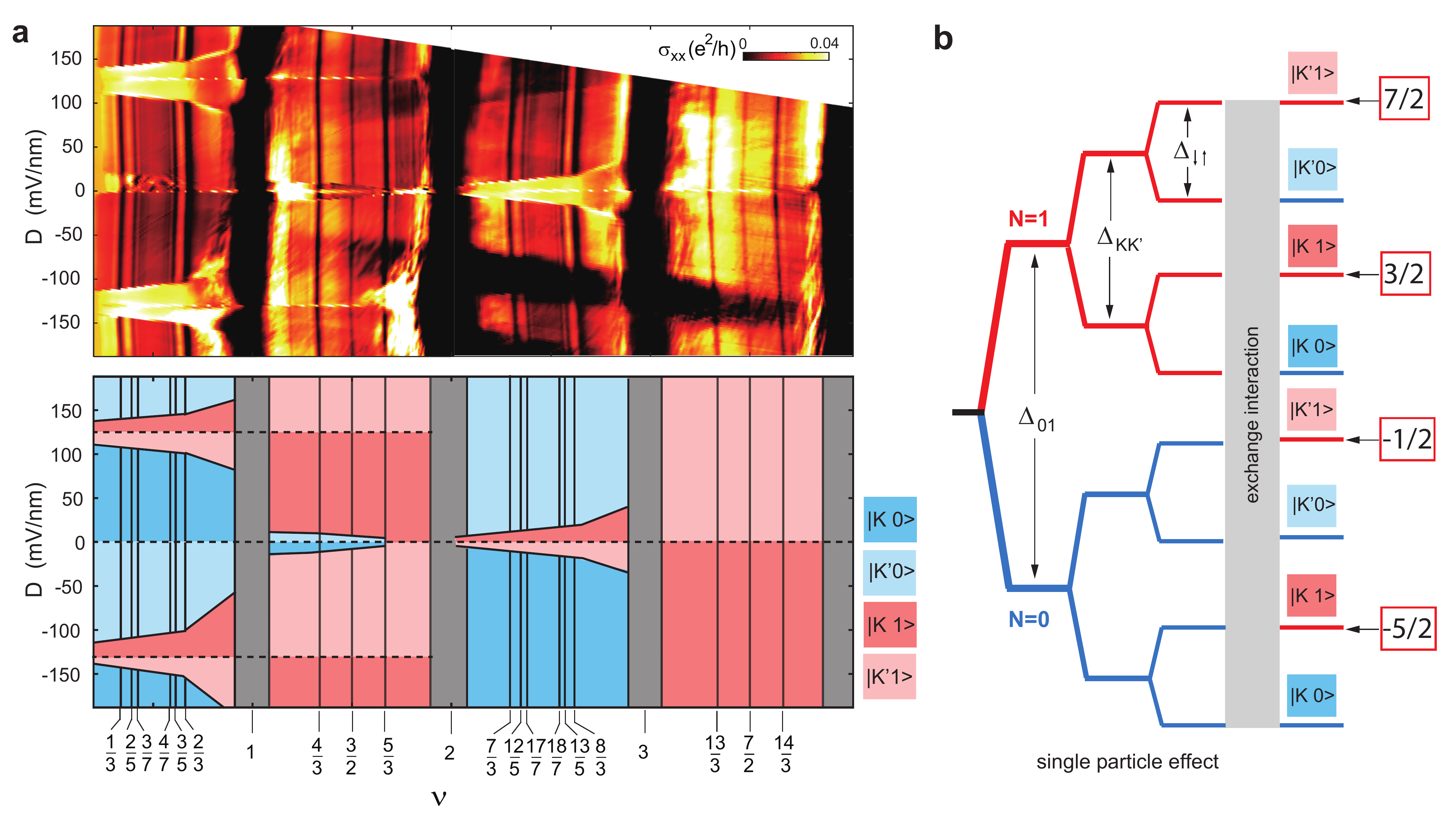}
\caption{\label{fig2}  (a) $\sigma_{xx}$ at $T = 0.2$ K and $B = 14.7$ T versus filling factor $\nu$ and displacement field $D$ for the LLL of bilayer graphene, $0 \leq \nu \leq 4$.  (b)  Schematic phase diagram labeling the ground sate order for the same filling fraction range as shown in (a). The blue (red) shaded area is occupied by broken-symmetry states with orbital index $0$ ($1$), whereas the dark and light color tones denote the two different valley-isospin (layer) polarizations. Dashed and solid black lines correspond to phase transitions between broken-symmetry states with different valley-isospin and orbital index, respectively. Vertical solid lines represent incompressible states observed in transport measurements.  (c) Single particle energy level diagram of the LLL (left) for constant $D$. Capacitance measurements ~\cite{Hunt.16} reveal a significant departure from the single-particle picture owing to the exchange interaction from completely filled and empty levels. }
\end{figure*}

\begin{figure*}
\includegraphics[width=1\linewidth]{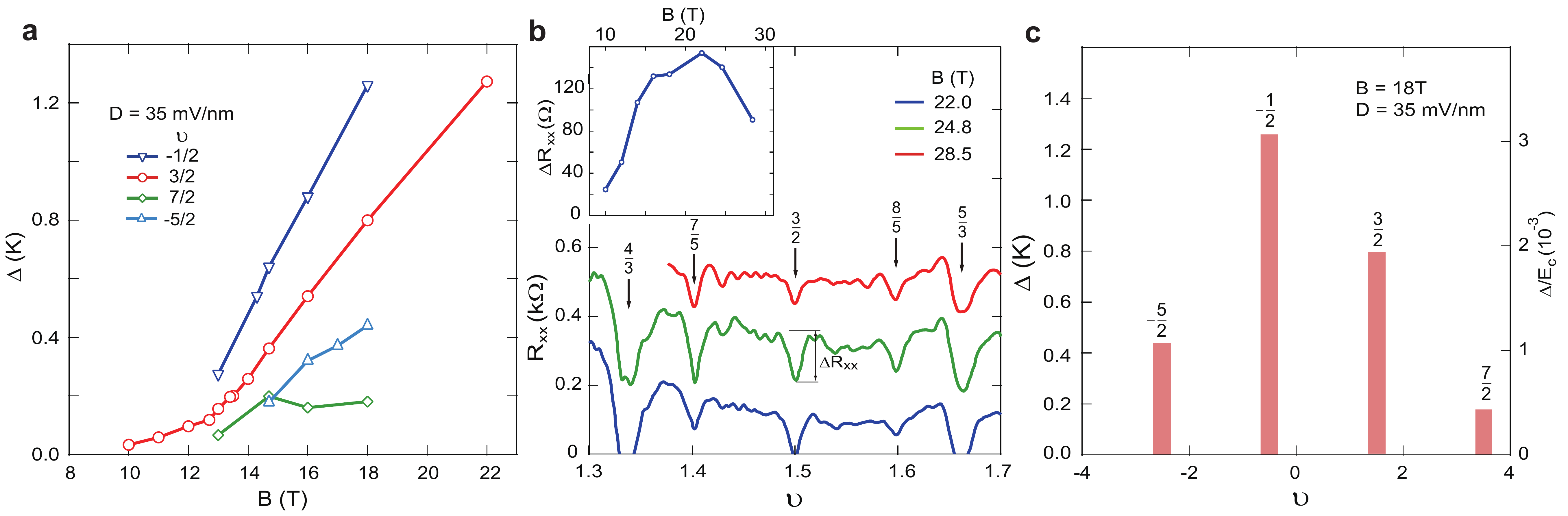}
\caption{\label{fig3}  (a) Energy gap $\Delta$ of the even denominator states as a function of magnetic field $B$, at $D = 35$ mV/nm.  (b) longitudinal resistance $R_{xx}$ versus $\nu$. Measurements are performed with different carrier density, $n = 7.7 \times 10^{11}$ (blue trace), $8.6 \times 10^{11}$ (green trace) and $9.9 \times 10^{11}$ cm$^{-2}$ (red trace), while sweeping $B$ so that the $\frac{3}{2}$ state appears at different $B$. Traces are offset for clarity. The inset shows the depth of the minimum, $\Delta R_{xx}$ versus $B$.   (c) $\Delta$ for all even denominator states at $B = 18$ T and $D = 35$ mV/nm. The right axis is normalized to the Coulomb energy $E_c = \frac{e^2}{\epsilon \ell_{B}}$. }
\end{figure*}

Fig.~1c shows the longitudinal conductance $\sigma_{xx}$ of the lowest Landau level (LLL) as a function of  filling fraction, $\nu$, and magnetic field. The data in this plot were acquired at fixed displacement field (see methods). 
The high device quality is evident by the ability to resolve all of the broken-symmetry IQHE states, as well as the numerous FQH states observed throughout the phase space ~\cite{Maher.14,Hunt.16,Zib.16}. Most remarkable is the  appearance of conductance minima, suggestive of several even-denominator states, occurring at filling fractions $-5/2$, $-1/2$, $3/2$, and $7/2$.  Based on recent understanding of how the eight-fold degeneracy of the LLL lifts at high field~\cite{Yacoby.14,Hunt.16}, we identify that these states appear only at half filling of the broken-symmetry states (BSSs) with orbital index $N=1$.   By contrast, no even denominator states appear within the $N=0$ BSSs ~\cite{Smet.15,Jain.17}.

Figs. 1e and 1d show traces of the  longitudinal resistance and transverse conductance  around $\frac{1}{2}$, and $\frac{3}{2}$ filling, acquired at fixed density $n$, and $D$, while varying $B$.  For $0 \leq \nu \leq 1$ (Fig.~1d) the wavefunction is characterized by orbital index $N=0$ ~\cite{Hunt.16}. A well developed series of FQHE states is observed, following the usual Jain sequence of composite Fermions (CFs) ~\cite{Jain.89}, with the highest resolvable CF state at $\nu=4/9$. At half filling, the longitudinal and Hall response both appear featureless.  Qualitatively this sequence closely resembles the FQHE hierarchy observed in the $N=0$ LLs in GaAs~\cite{Wil.87}. In stark contrast is the behaviour between $1 \leq \nu \leq 2$, corresponding to a BSS with orbital index $N=1$  (Fig. 1e).  Fewer FQHE states are resolved, with only the $n/3$ CF states appearing fully developed.  A robust incompressible state is clearly present at $\nu = \frac{3}{2}$, marked by zero longitudinal resistance simultaneous with a quantized Hall plateau.  The qualitative similarity to the $N=1$ states observed in GaAs~\cite{Wil.87,Pan.99} and more recently in ZnO ~\cite{Falson.15}, both in terms of the even-denominator state appearing at half filling, and its strength relative to the FQHE states away from half filling, highlights the important role played by the orbital wavefunction in determining the interaction driven behavior.  Moreover, the $N=1$ orbital selection rule suggests that the even denominator FQHE observed here may be the same MR Pfaffian ground state in GaAs ~\cite{Wil.87,Pan.99}.

\begin{figure*}
\includegraphics[width=0.75\linewidth]{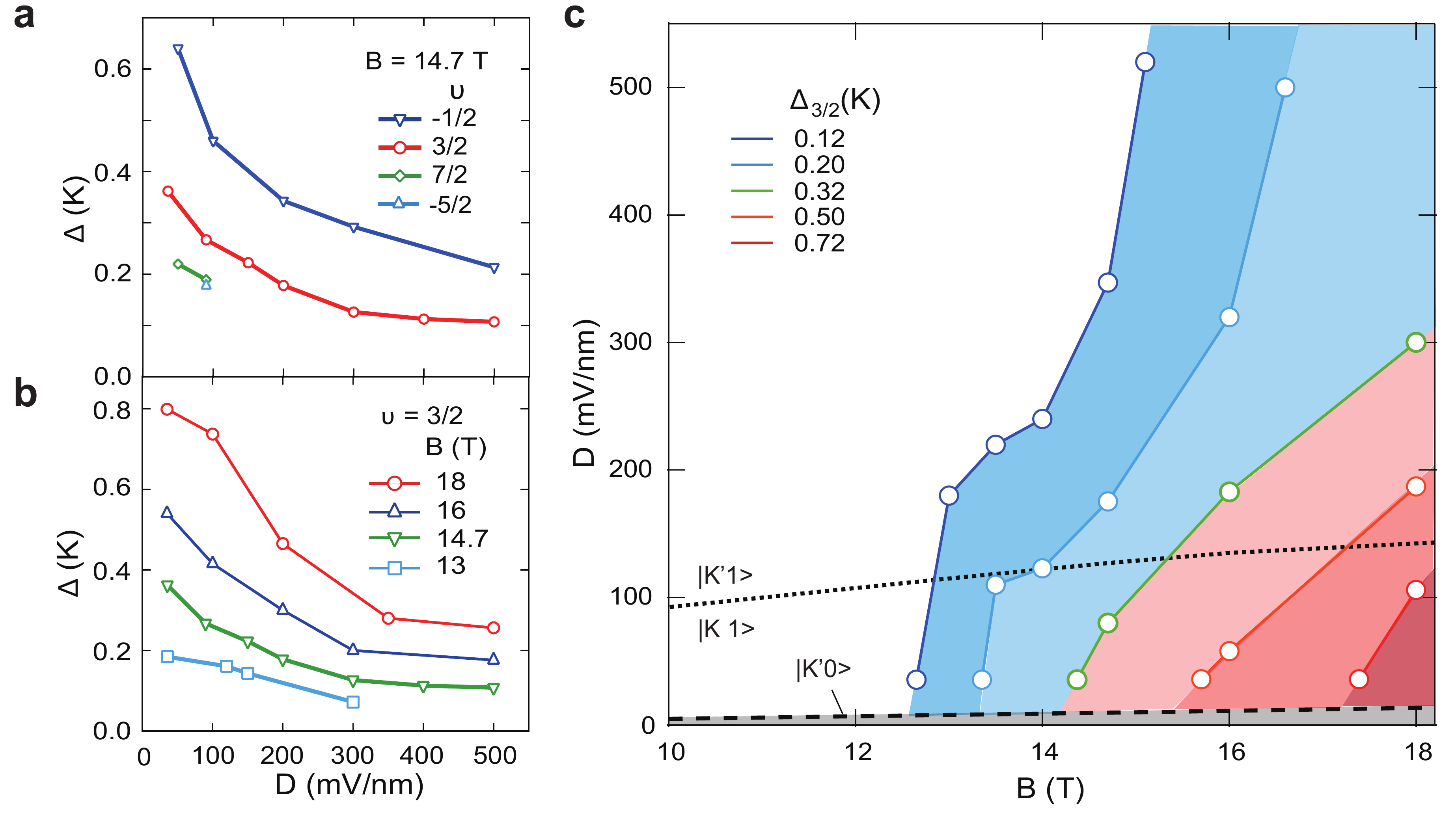}
\caption{\label{fig4}  (a) upper panel, $\Delta$ as a function of $D$ for $\nu = -\frac{1}{2}$, $\frac{3}{2}$, $-\frac{5}{2}$ and $\frac{7}{2}$ at $B = 14.7$ T. (b) lower panel, $\Delta$ versus $D$ for $\nu = \frac{3}{2}$ at different $B$. (c) Contour line for $\Delta_{\frac{3}{2}}$ as a function of $B$ and $D$. The valley polarization transition at non-zero $D$ is shown as the black dotted line and the orbital polarization transition as the black dashed line. The even-denominator state disappears below the orbital transition (gray shaded area), where the orbital $0$ wavefunction is stabilized by screening interactions. }
\end{figure*}

Fig.~\ref{fig2}a plots $\sigma_{xx}$ versus $D$ and $\nu$ for $0 \leq \nu \leq 4$, the electron half of the LLL (similar analysis of the valence band states is shown in the SI). A schematic phase diagram of the ordering associated with each state is shown in Fig. 2b.  Identifying the precise ground state  for each BSS is difficult since interaction effects can reorder the LLs expected from a simple single particle description of degeneracy lifting (Fig.~2c).  Horizontal features in Fig. 2a (local maxima in $\sigma_{xx}$) can be used to identify phase transitions  between different orbital  and valley polarizations ~\cite{Hunt.16}, but cannot unambiguously distinguish between the  different possible orders.  Measurement of the layer polarization by capacitance spectroscopy was recently used to map the valley and orbital components within the LLL; however, the spin ordering remains uncertain ~\cite{Hunt.16}.  We therefore label the phase diagram with a reduced 2-component wavefunction $ \left | \zeta N \right \rangle $, where $\zeta = K$, $K'$ is the valley isospin, and $N=0$ and $1$ the Landau orbital, without identifying the spin component.  

Qualitatively we find that even denominator states are insensitive to the changes in the valley ordering, persisting smoothly across $D$-induced transition boundaries.  For example, the $7/2$ state is well developed over both the $K'$ ($D>0$) and $K$ ($D<0$) regions. Likewise the $3/2$ state remains stable through four different valley orderings as $D$ is varied. In this case we note that the $3/2$ state disappears near $D=0$ but that this corresponds to a transition to a $N=0$ orbital, consistent with the same selection rule identified above.  That the even denominator state is sensitive to orbital structure but insensitive to valley order is consistent with theoretical treatment of the Pfaffian in BLG ~\cite{Chak.11,Pap.11}.  However, it is curious that the state remains stable near the transition boundary, where the effects of LL mixing are expected to be most prominent ~\cite{Bishara.09,Rezayi.17} and typically assumed to play a destructive role. We also note that no FQH state appears to form at $1/2$ and $5/2$ filling, even within regions where a $D$-induced orbital transition to $N=1$ is realized.

In order to fully characterize the tunability of the even denominator states, we examine the energy gap by variable-temperature activation measurements. Up to $B = 22$ T the  gap, $\Delta$, of all the even denominator states grows monotonically with $B$ (Fig. 3a).  This is consistent with an enhanced Coulomb interaction, $E_c = \frac{e^2}{\epsilon \ell_{B}}$, where $e$ is the electron charge, $\epsilon$ is the dielectric constant and $\ell_{B} \approx 26$ nm $/\sqrt{B}$ the magnetic length. However, it has also been proposed that since the $N=1$ wavefunction in BLG is a mixture of conventional Landau orbital $0$ and $1$ wavefunctions~\cite{Pap.11,Chak.11,Hunt.16}, where the relative $0$ orbital contribution  grows with increasing $B$, the even denominator gap may be suppressed beyond some critical field~\cite{Pap.11,Chak.11}. While we do not have access to activation gap measurements above $B=22~$ T,  we observe qualitatively that the depth of the resistance minimum does indeed  diminish as the field approaches $30$~T (Fig. 3b). This is suggestive of a B-field induced transition to a gapless state.


Interestingly, among the four even denominator states, the size of energy gap demonstrates a hierarchy, as shown in Fig.~\ref{fig3}c, where $\Delta$ becomes weaker  approaching  the edge of the LLL ($\Delta_{-\frac{1}{2}}$ is the strongest and $\Delta_{\frac{7}{2}}$ the weakest). Several effects may contribute to this observation, including variation in level mixing, electron interaction strength, and disorder, all of which may be filling fraction dependent within the LLL.  Further experimental and theoretical work will be necessary to fully resolve this trend.

Fig.~4a shows the effect of displacement field on the variation of $\Delta$. All of the even denominator states are found to be strongly suppressed with increasing $D$. For example, at fixed $B = 14.7$ T and $D = 500$ mV/nm, both $\Delta_{-\frac{1}{2}}$ and  $\Delta_{\frac{3}{2}}$ show reduction of more than a factor of $2$ compared to $D = 35$ mV/nm. The $D$-induced gap variation is significantly larger than theoretical predictions~\cite{Chak.11}, and the origin of this surprising result is not known. However, we note that $D$ dependence appears to vary with $B$ (Fig.~4b).  This behavior is summarized in the experimentally measured $B - D$ phase diagram for the $3/2$ state gap, shown in Fig.~4c. Also shown in this diagram are  the $D$-field values for the valley (black dotted line) and orbital (black dashed line) transitions. Within our data resolution there is no discernible dependence on the valley order.  $\Delta_{\frac{3}{2}}$ is too small to measure in the white area, and is absent all together in the gray shaded area, where the ground state wavefunction has orbital index $N=0$.

An important question remains: what is the ground state wavefunction of the even denominator state? A frequently considered alternative to the MR Pfaffian is the  Halperin (331) ~\cite{Hal.83}, an Abelian ground state resulting from a spin singlet (or possibly also valley-isospin singlet) pairing of electrons.  This state however was theoretically found to be unstable in BLG, except for narrowly constrained conditions~\cite{Pap.14}, and therefore unlikely to explain the wide ranging phase space that we observe.  Experimentally we note that without a direct probe of spin order, we can not definitely rule out the possibility of a spin-singlet.  We consider a valley singlet pairing unlikely since we observe the even-denominator state to persist to high $D$, where the state is assumed to be fully valley polarized ~\cite{Yacoby.10,Yacoby.14,Maher.14,Hunt.16}.  In addition to the Pfaffian, similar non-abelian ground states such as its particle-hole conjugate, the anti-Pfaffian~\cite{Levin.07}, may be possible, depending on the strength of LL  mixing~\cite{Rezayi.17}.  Interestingly, the ability to tune the degree of LL mixing in graphene could make it possible to dynamically transition between the Pfaffian and anti-Pfaffian, which are topologically distinct ground states.

The observation of a robust and dynamically tunable even denominator FQHE state in BLG provides a rich new platform in which to probe the nature of the Pfaffian ground state. The ability to reach a regime where the transport gap exceeds $1$K makes the state accessible to a wider range of experimental probes than previously possible in GaAs.  Moreover the use of thin hBN dielectrics combined with the 2-dimensional (2D) nature of the BLG crystal  may enable the electrostatic control necessary to investigate the presumed non-Abelian statistics through new interferometry experiments ~\cite{Bishara.08}

\subsection{Methods}
The device geometry, shown in Fig.~1a, includes both top and bottom gate electrodes made of few-layer graphite. The heterostructure is assembled using the van der Waals transfer technique ~\cite{Lei.13}, so that the BLG flake as well as the top and bottom hBN dielectrics are free from contamination during the stacking and subsequent fabrication processes. Graphene voltage leads extend outside of the local graphite gate, and can be tuned to high density by the the doped Si substrate covered by a $285$-nm-thick SiO$_2$ layer. Four-terminal electrical measurements were used for transport characterization.

\subsection{acknowledgments}
The authors thank Z. Papic and A. Young for helpful discussion. This work was supported by the National Science Foundation (DMR-1507788). C.R.D acknowledges partial support by the David and Lucille Packard Foundation. T.C is supported by INDEX. A portion of this work was performed at the National High Magnetic Field Laboratory, which is supported by National Science Foundation Cooperative Agreement No. DMR-1157490 and the State of Florida.

\end{document}